\documentclass{gMOS2e}

\usepackage{epstopdf}
\usepackage{subfigure}

\usepackage{amsmath}
\usepackage{graphicx}
\usepackage[caption=false]{subfig}

\begin{document}
\title{Disordered vortex matter out of equilibrium: a Langevin molecular dynamics study}

\author{
\name{Hiba Assi\textsuperscript{\,a}, Harshwardhan Chaturvedi\textsuperscript{\,a},
Ulrich Dobramysl\textsuperscript{\,b}, Michel Pleimling\textsuperscript{\,a,c}
$^{\ast}$\thanks{$^\ast$Corresponding author. Email: pleim@vt.edu} and
Uwe C. T\"auber\textsuperscript{\,a}}
\affil{\textsuperscript{a}Department of Physics (MC 0435), 850 West Campus Drive, 
        Virginia Tech, Blacksburg, VA 24061, USA;
\textsuperscript{b}Mathematical Institute, University of Oxford, 
        Andrew Wiles Building, Radcliffe Observatory Quarter, Woodstock Road, 
        Oxford OX2 6GG, UK;
\textsuperscript{c}Academy of Integrated Science (MC 0405),
300 Turner Street NW, Virginia Tech, Blacksburg, VA 24061, USA}
}
\maketitle

\begin{abstract}
We discuss the use of Langevin molecular dynamics in the investigation of the non-equilibrium 
properties of disordered vortex matter. Our special focus is set on values of system parameters 
that are realistic for disordered high-$T_c$ superconductors such as YBCO. Using a discretized 
elastic line model, we study different aspects of vortices far from thermal equilibrium. On the 
one hand we investigate steady-state properties of driven magnetic flux lines in a disordered
environment, namely the current-voltage characteristics, the gyration radius, and the pinning 
time statistics. On the other hand we study the complex relaxation processes and glassy-like 
dynamics that emerge in type-II superconductors due to the intricate competition between the 
long-range vortex-vortex repulsion and flux pinning due to randomly placed point defects. To 
this end we consider different types of sudden perturbations: temperature, magnetic field, 
and external current quenches.

\end{abstract}


\section{Introduction}
\label{sec:introduction}

Langevin equations are stochastic differential equations that have seen widespread applications
in many areas of physics \cite{Coffey_Kalmykov:12}. They form a sound effective description in 
situations where a strict separation of time scales prevails, i.e., where one has both fast and slow 
degrees of freedom. Langevin equations then provide equations of motion for the relevant slow 
modes, whereas the effects of the fast degrees (e.g., the environment) are incorporated via
stochastic forcing.

Since the initial use of a Langevin equation for the description of a Brownian particle
\cite{Langevin:08}, many other problems have been tackled using stochastic differential equations. 
Obvious problems can be found in biological settings where large 'particles' are moving in an 
aqueous environment. But there are also important problems in condensed matter and material 
physics that can be successfully addressed using Langevin equations. Well-known examples 
include domain walls in magnetic systems, fluctuating interfaces, or flux lines in type-II 
superconductors.

In this paper we discuss the use of Langevin equations for the study of interacting vortices in a 
disordered environment. Our main focus is on high-$T_c$ superconductors \cite{Blatter1994} 
where glassy states are realized at low temperatures due to the competition of long-range 
vortex-vortex interaction and short-range defect pinning. More specifically, we will discuss in some 
detail the properties of these systems when they are forced far away from thermal equilibrium. By 
employing the methods of Langevin molecular dynamics, we numerically solve the set of coupled 
stochastic differential equations describing our systems. In particular, our model systems are well 
suited for the application of this method due to the separation into many slow degrees of freedom 
embedded in an environment dominated by fast flluctuating thermal degrees of freedom.

In principle, a system can be prepared 'far from equilibrium' in two different ways. In the first case
the system is subjected to an external drive that yields a non-equilibrium steady state.
Non-equilibrium steady states have been a focus of many studies in statistical physics that aim at 
finding a comprehensive theoretical framework allowing to determine the stationary probability 
density and probability current distribution. Whereas some progress has been made in certain 
instances, a comprehensive theory remains elusive. Non-equilibrium steady states are also of 
interest in the context of vortex matter as they emerge naturally when applying a driving force to 
the system. 

Another way to induce a system out of equilibrium is to force it away from a stationary state 
through a sudden change of at least one of the system parameters \cite{Struik1978}. This sudden 
change, often called quench, can be realized in different ways. Very common is a temperature 
quench, but other types of quenches (magnetic field or vortex density as well as external current
or drive quenches) are also possible and will be discussed in this paper. Once the system has been 
brought out of the initial steady state, it will relax towards a new stationary state. This relaxation 
process can be very long-lasting in systems with slow or glassy-like dynamics. In some instances 
this relaxation process manifests itself through characteristic dynamical scaling behavior 
\cite{Henkel2010}. In other cases relaxation is a very involved process, due to the competition 
between different physical interactions or other effects, and time-dependent quantities may 
exhibit a very rich and complex behavior. The intricate slow relaxation processes of interacting 
magnetic vortices in disordered type-II superconductors deep inside the low-temperature Bragg
glass phase are of this second type.

This paper is meant to provide a didactic introduction to the investigation of non-equilibrium
properties of flux lines in a disordered environment using Langevin molecular dynamics. After 
introducing the basics of Langevin dynamics in the next section, we focus in section 3 on the 
study of interacting vortex matter using Langevin molecular dynamics. We first recall the effective
description through an elastic line model before describing the implementation of Langevin 
molecular dynamics for this system. Section 4 presents a range of results, both for driven vortices
in the steady state as well as for non-equilibrium relaxation processes that result from a quench 
of one of the external system control parameters. We conclude in section 5.

\section{Langevin dynamics}

Before we employ Langevin molecular dynamics for the investigation of the non-equilibrium
properties of vortices in a disordered environment in the next section, it is useful to first illustrate 
the concept of a Langevin equation through simpler examples. For that reason we will briefly
recall the case of a Brownian particle, followed be a general discussion of the Langevin equation 
of a system characterized by a Hamiltonian and the associated Boltzmann-Gibbs equilibrium 
probability distribution.

\subsection{Brownian particle}

The paradigmatic example of a Langevin equation is of course the stochastic equation that 
describes a particle immersed in a fluid. This particle is continuously exposed to collisions with the 
molecules of the fluid. Assuming that there is no net flow, the molecules hit the particle with the 
same rate from all sides, i.e., on average the force on the particle due to the fluid molecules is
zero. In a probabilistic description this can be modeled by a stochastic force with zero mean. As 
the particle moves through the fluid, a hydrodynamic drag force sets in as a consequence of the
random collisions with the fluid molecules that impedes the motion of the particle. Taking these 
two effects into account, the equation of motion governing the particle is given by the Langevin 
equation
\begin{equation}
\label{eq:Brownian}
m \frac{d{\bf v}(t)}{dt} = - \eta {\bf v}(t) + {\bf f}(t)~.
\end{equation}
Here ${\bf v}(t)$ and $m$ are the instantaneous velocity and the mass of the particle, 
respectively, $\eta$ is the friction coefficient, and {\bf f}(t) represents the stochastic force.

Before discussing the noise term further, let us stress that a stochastic description like equation 
(\ref{eq:Brownian}) assumes a separation of time scales between slow degrees of freedom (in 
this case the particle) and fast degrees of freedom (the fluid molecules). Usually the fast 
degrees of freedom represent the environment in which the slow degrees evolve.

For the stochastic noise the following two assumptions are usually made: (1) the collisions of the
fluid molecules with the particle are uncorrelated in time, and (2) the random forces are drawn
from a Gaussian probability distribution with zero mean: 
\begin{eqnarray}
\langle {\bf f}(t) \rangle & = & 0~,\\
\langle {\textrm f}_\nu(t) \, {\textrm f}_\mu(t') \rangle & = & \gamma \delta_{\nu \mu} 
\delta(t-t')~, 
\end{eqnarray}
where $\nu$ and $\mu$ denote Cartesian coordinate indices and $\gamma$ is the strength of 
the noise. While these assumptions are perfectly sound for a Brownian particle, one should 
make sure for each situation whether they are indeed meaningful for the problem at hand.

The noise term $\gamma$, which measures the strength of the fluctuations, and the friction 
coefficient $\eta$, which describes dissipation, are related through the fluctuation-dissipation 
theorem (called Einstein relation in the present context):
\begin{equation}
\gamma = 2 k_B T \eta~,
\end{equation}
where $T$ is the temperature of the environment and $k_B$ is Boltzmann's constant.

If the inertia time is larger than $m/\eta$, inertial effects can be disregarded, 
and the particle's motion is then 
governed by the overdamped Langevin equation
\begin{equation}
\label{eq:Brownian_dynamics}
\eta \frac{d{\bf r}(t)}{dt} = {\bf F}({\bf r}(t)) + {\bf f}(t)~,
\end{equation}
where we have added a conservative force ${\bf F}({\bf r}) = - \nabla V({\bf r})$ originating
from a potential $V({\bf r})$. Equation (\ref{eq:Brownian_dynamics}) is in fact the starting
point for Brownian dynamics as well as for Langevin molecular dynamics used in the following. 
Note that imposing the fluctuation-dissipation theorem ensures that this stochastic differential 
equation correctly describes the equilibrium properties of our particle.

\subsection{Classical systems governed by a Hamiltonian}

The overdamped Langevin equation (\ref{eq:Brownian_dynamics}) is readily generalized to 
situations with $N$ degrees of freedom (spatial coordinates of the particles) governed by a 
Hamiltonian $H[\left\{ {\bf r} \right\}]$ that is a functional of the position vectors ${\bf r}$
of the different slow degrees of freedom. In that case one has the following set of coupled 
stochastic differential equations
\begin{equation}
\label{eq:lmd}
\eta \, \frac{\partial {\bf r}_i(t)}{\partial t} = - \frac{\delta H[{\bf r}_i(t)]}{\delta {\bf r}_i(t)} 
+ {\bf f}_i(t)~,
\end{equation}
where $i=1,\ldots,N$ labels the different particles. As before, the effects of the fast 
degrees of freedom coming from the environment are captured by the stochastic force 
${\bf f}_i(t)$ acting on particle $i$. These stochastic forces have again zero mean, 
$\langle {\bf f}_i(t) \rangle = 0$, and are taken from a Gaussian distribution, thereby 
fulfilling the Einstein relation (in $d$ spatial dimensions)
\begin{equation}
\label{eq:FDT}
\langle {\bf f}_i(t) \cdot {\bf f}_j(t') \rangle = 2 d \eta k_B T \delta_{ij} \delta(t - t')~.
\end{equation}
This set of equations guarantees that in equilibrium the system samples configurations with 
the correct canonical equilibrium probability distribution, $\sim e^{-H / k_B T}$. However, out 
of equilibrium, which is our primary interest in the following, it is important to note that different 
microscopic dynamics might yield different results. This is even true in cases where the same 
probability distribution is ultimately obtained in equilibrium. A detailed comparison between 
different implementations of the system's dynamics might therefore be in order if one wishes to 
ascertain that out-of-equilibrium results are not merely artefacts of the dynamics (see, e.g.,
Ref.~\cite{Dobramysl2013}).

\section{Langevin molecular dynamics for interacting vortex matter}

\subsection{Elastic line model}

We consider in the following $N$ interacting elastic lines in a disordered environment. Having 
in mind magnetic flux lines in high-$T_c$ superconductors, we remark that this situation 
corresponds to the extreme London limit with the superconducting coherence length much 
smaller than the London penetration depth \cite{Nelson_Vinokur:93}. Vortex lines are then 
described by their trajectories ${\bf r}_i(z)$, where $z$ denotes the direction of the applied 
external magnetic field, and the two-dimensional vector ${\bf r}_i$ indicates the $xy$ position 
of line $i$ at height $z$. As an immediate consequence of this description, magnetic flux lines in 
this model cannot form loops or overhangs, since ${\bf r}_i(z)$ has to be unique. This is a 
reasonable assumption, as long as the vortex energy per length (i.e,. the elastic line tension) 
remains large compared to the energy scale of thermal fluctuations.

The effective Hamiltonian of this system is written as a functional of the vortex line trajectories 
with an extent of $L$ in the $z$ direction and consists of three competing terms: the elastic line
energy, the external potential due to disordered pinning sites, and the repulsive vortex-vortex 
interaction:
\begin{eqnarray}
   H_N[{\bf r}_i] &=& \frac{\tilde{\epsilon}_1}{2} \sum_{i=1}^N \int_0^L \bigg\arrowvert
   \frac{d{\bf r}_i(z)}{dz} \bigg\arrowvert^2 \, dz 
   + \sum_{i=1}^N \int_0^L V_D\bigl( {\bf r}_i(z) \bigr) \, dz \nonumber \\
   && + \frac{1}{2} \sum_{i \ne j} \int_0^L V\bigl(|{\bf r}_i(z)-{\bf r}_j(z)| 
   \bigr) \, dz \ .
\label{hamilt}
\end{eqnarray}
Consistent with the extreme London limit, the repulsive vortex-vortex pair interaction is set
purely in-plane between different flux line elements. Below we will provide expressions for the 
attractive pinning potential and repulsive interaction appropriate for high-$T_c$ superconductors.

In order to employ a Langevin molecular dynamics algorithm to simulate the vortex line dynamics, 
we discretize the system into layers along the $z$ axis. Forces acting on the flux line vertices can 
then be derived from the properly discretized version of the Hamiltonian (\ref{hamilt}). We 
subsequently proceed to numerically solve the set of coupled overdamped Langevin equations
\begin{equation}
\eta \, \frac{\partial {\bf r}_i(t,z)}{\partial t} = - \frac{\delta H_N[{\bf r}_i(t,z)]}{\delta {\bf r}_i(t,z)} 
+ {\bf f}_i(t,z) + {\bf F}_d~.
\label{langevin_sc}
\end{equation}
As discussed in the previous section, the fast, microscopic degrees of freedom of the surrounding 
medium are captured by thermal stochastic forces ${\bf f}_i(t,z)$, modeled as uncorrelated 
Gaussian white noise with vanishing mean and fulfilling the Einstein relations (\ref{eq:FDT}) in 
$d=2$ dimensions. This guarantees that the system relaxes to thermal equilibrium with a 
canonical probability distribution in absence of an external driving force ${\bf F}_d$. In type-II 
superconductors such an external driving force stems from applied external currents via the 
Lorentz force.

The time integration is performed via simple discretization of equation (\ref{langevin_sc}) 
\cite{Brass1989}. In Ref.~\cite{Dobramysl2013} we also considered the situation of small
inertia times. The Langevin equation then acquires an additional inertial term and the time
stepping algorithm becomes slightly more complicated \cite{Brunger1984}.

\subsection{Implementation for disordered type-II superconductors}

High-$T_c$ superconducting materials are layered compounds and highly anisotropic: the lattice 
constant in the crystallographic $c$ direction is much larger than the in-plane ones along the  $a$ 
and $b$ directions; correspondingly, the effective charge carrier masses $M_{ab}$ and $M_c$
differ as well for the different directions. Henceforth we always assume that the magnetic field is
aligned with the material's crystallographic $c$ direction, and the material properties discussed
below are given for this configuration and assigned the in-plane index $ab$. When discretizing the 
system into layers along the $z$ axis we choose the crystallographic $c$ axis unit cell size $c_0$
as layer spacing \cite{Das2003, Bullard2008}.

This anisotropy also shows up in the expressions for the model parameters when using the
effective elastic line model (\ref{hamilt}). The elastic line stiffness or local tilt modulus is given by
$\tilde{\epsilon}_1 \approx \tilde{\Gamma}^{-2} \epsilon_0 \ln \left( \lambda_{ab}/\xi_{ab} \right)$.
Here $\lambda_{ab}$ is the London penetration depth and $\xi_{ab}$ is the coherence length, 
both in the $ab$ plane, whereas $\tilde{\Gamma}^{-1} = M_{ab}/M_c$ denotes the effective mass 
ratio. The energy per length $\epsilon_0$ is given by $\epsilon_0 = (\phi_0 / 4 \pi 
\lambda_{ab})^2$, with the magnetic flux quantum $\phi_0 = h c / 2 e$. Finally the 
Bardeen-Stephen viscous drag coefficient $\eta$ in the Langevin equation (\ref{langevin_sc}) is 
$\eta = \phi_0^2 / 2 \pi \rho_n c^2 \xi_{ab}^2$, where $\rho_n$ represents the normal-state 
resistivity \cite{Bardeen1965}.

For our implementation of the Langevin molecular dynamics method for disordered type-II 
superconductors \cite{Dobramysl2013, Dobramysl2014, Assi2015, Chaturvedi2015} we also 
require expressions for the vortex-vortex interaction as well as for the attractive pinning potential.
The repulsive in-plane vortex-vortex interaction is given by
\begin{equation}
V(r) = 2 \epsilon_0 K_0 (r/\lambda_{ab})~, 
\end{equation}
where $K_0$ denotes the zeroth-order modified Bessel function. This is essentially a logarithmic
repulsion that is exponentially screened at the scale $\lambda_{ab}$. In order to avoid artefacts
due to periodic boundary conditions in directions perpendicular to $z$, we cut off the interaction
at $5 \lambda_{ab}$. Pinning sites are modeled by randomly distributed smooth potential wells of 
the form
\begin{equation}
V_\alpha({\bf r},t) = - \frac{b_0}{2} \, p \, \delta(z - z_\alpha) \left[
1 - \tanh \left(5 \frac{\left| {\bf r} - {\bf r}_\alpha \right| - b_0}{b_0} \right) \right]~,
\end{equation}
where $p > 0$ is the pinning potential strength, and $z_\alpha$ and ${\bf r}_\alpha$ denote the 
$z$ position and in-plane location of pinning site $\alpha$, respectively. The pinning potential width 
$b_0$ will be our unit of length, whereas energies will be measured in units of $\epsilon_0 b_0$. 
The full pinning potential $V_D({\bf r},t)$ is then obtained by summing over all $N_D$ localized 
pinning sites in the system: 
\begin{equation}
V_D({\bf r},t) = \sum\limits_{\alpha=1}^{N_D} V_\alpha({\bf r},t)~.
\end{equation}
For the case of random point defects, which is the only case considered in this paper, these pinning 
centers are randomly distributed and chosen independently for each layer. If one were to instead 
consider columnar defects aligned parallel to the magnetic field along the $z$ direction, then one 
would repeat the same spatial distribution pattern for each layer \cite{Dobramysl2013, Assi2015, 
Pleimling2015}.

The numbers we use for our model parameters are consistent with the actual values for YBCO 
\cite{Blatter1994}. We set the pinning potential width $b_0 = 35 \AA$ and also choose $c_0 = b_0$. 
The superconducting coherence length is then given by $\xi_{ab} = 0.3\,b_0$, whereas the London 
penetration depth is $\lambda_{ab} = 34\,b_0$. The effective mass anisotropy ratio is
$\tilde{\Gamma}^{-1} = 0.2$ and the normal-state resistivity near $T_c$ becomes $\rho_n \approx 
500 ~\mu \Omega~\mbox{cm}$. With this, the numerical values of all model parameters can be 
calculated. It then also follows that our fundamental temporal unit is $t_0 = \eta b_0 / \epsilon_0 
\approx 18$ ps. In the following we shall measure all time intervals in units of $t_0$.

\subsection{Quantities of interest}

A large range of quantities of interest are at one's disposal when studying the out-of-equilibrium
properties of vortex matter. Some of these quantities are especially useful to characterize the
steady-state properties in the presence of an external driving force, others allow us to capture the 
complex relaxation processes in situations where the system is initialized far from a steady state.

A typical experimentally accessible quantity is the current-voltage (I-V) characteristics. From a 
Langevin molecular dynamics study an equivalent description is obtained from a graph relating the 
driving force ${\bf F}_d$ to the mean vortex velocity ${\bf v}$. Indeed, from Faraday's law the 
mean vortex velocity, which results from the velocities of each line element in direction of the 
driving force, is related to an induced electric field ${\bf E} = {\bf B} \times {\bf v}/c$ and therefore
to a voltage drop across the sample. A relation between the driving force ${\bf F}_d$ and an 
applied external current density ${\bf j}$ follows from the Lorentz force:  
${\bf F}_d = {\bf j} \times \phi_0 {\bf B}/B$.

The thermal spatial fluctuations along the vortex lines can be captured by the radius of gyration. 
Defined as $r_g = \sqrt{  \langle ( {\bf r}_i(z) - \langle {\bf r}_i \rangle )^2 \rangle }$, this quantity 
is in fact nothing else than the average root mean-square displacement from the mean lateral 
positions of the lines. Here, $\langle \cdots \rangle$ indicates an average over all line elements of 
line $i$ as well as an average over all $N$ lines and different realizations of the disorder and the 
noise.

A third way to explore steady-state properties of driven vortex matter is to study the pinning time 
statistics. This is done by calculating the distribution of flux line element dwelling times at defect 
sites. The dwelling time at pins is obtained by monitoring the positions of flux line elements and 
recording the instances where they enter and leave the attractive pinning centers.

Non-equilibrium relaxation processes are best studied by preparing a system in a specific state 
before suddenly changing a system parameter \cite{Henkel2010}. This sudden change, often 
referred to as a quench, can be realized in many ways, depending on the physical system under 
investigation. For disordered vortex matter the following three types of quenches can be realized
experimentally: temperature quenches, magnetic field (or vortex density) quenches, and external
current (i.e., driving force) quenches. After such a quench the system ends up in a state far from 
stationarity. The resulting relaxation processes can then be monitored through various quantities.

Many previous studies of glasses (structural glasses or spin glasses) as well as of systems with 
glassy-like dynamics have revealed that relaxation and related aging processes are best 
investigated through two-time quantities \cite{Henkel2010}. Indeed, as a system after the quench
is far from stationarity, time translation invariance is broken and the properties of the system
change with the time elapsed since the quench. As a result an observable that depends on two 
times $s$ and $t > s$ is not only a function of the time difference $t-s$ but depends in more 
complicated ways on these two times.

In the context of interacting vortices in a disordered environment an impressive list of two-time 
quantities has been discussed \cite{BCD1, BCD2, BCI, IBKC, Pleimling2011, Dobramysl2013, 
Assi2015, Chaturvedi2015}. Some of these quantities contain information about local thermal 
fluctuations, whereas others provide insight into the time evolution of the global structure of the 
flux line configuration. We refrain from discussing all these quantities, but instead only introduce 
those used and displayed in the following concise overview:
The height-height autocorrelation function (sometimes simply called the roughness) probes the 
local thermal fluctuations of vortices about their mean lateral position. It is defined as 
\begin{equation}
C(t,s) = \big\langle \big( {\bf r}_i(t,z) - \langle {\bf r}_i(t) \rangle \big) 
\big( {\bf r}_i(s,z) - \langle {\bf r}_i(s) \rangle \big) \big\rangle~,
\end{equation}
where the averages are again taken over the flux line elements of all lines as well as noise and
disorder realizations. The average square distance between a vortex line element's position at 
the two times $s$ and $t$ is measured by the two-time mean square displacement
\begin{equation}
B(t,s) = \big\langle \big( {\bf r}_i(t,z) - {\bf r}_i(s,z) \big)^2 \big\rangle~.
\end{equation}
This quantity contains information on the decay and formation of global structures.

\section{Disordered vortex matter out of equilibrium}

We illustrate the rich physics of interacting magnetic vortices in disordered type-II superconductors 
that is governed by competing energy and hence length and time scales, through various characteristic
examples that address both fluctuations in non-equilibrium steady states \cite{Dobramysl2013, 
Chaturvedi2015, Dobramysl2014} as well as complex out-of-equilibrium relaxation scenarios initiated 
from very different starting configurations \cite{Dobramysl2013, Assi2015, Chaturvedi2015}. In this 
brief overview, we focus on systems with comparatively weak magnetic fields and hence low vortex 
density, and at low temperatures. Consequently the equilibrium stationary configurations in the 
presence of attractive point-like pinning centers likely reside in the disorder-dominated Bragg glass 
thermodynamic phase, wherein spatial positional order is disrupted at long length scales. At the cost of 
elastic energy, the flux lines attempt to accommodate as many localized defect sites as possible, which 
renders their trajectories through the sample quite rough \cite{Nattermann2000}.

\subsection{Driven vortex lines in disordered environments}

\begin{figure}[t]
\centering \includegraphics[width=0.44\columnwidth]{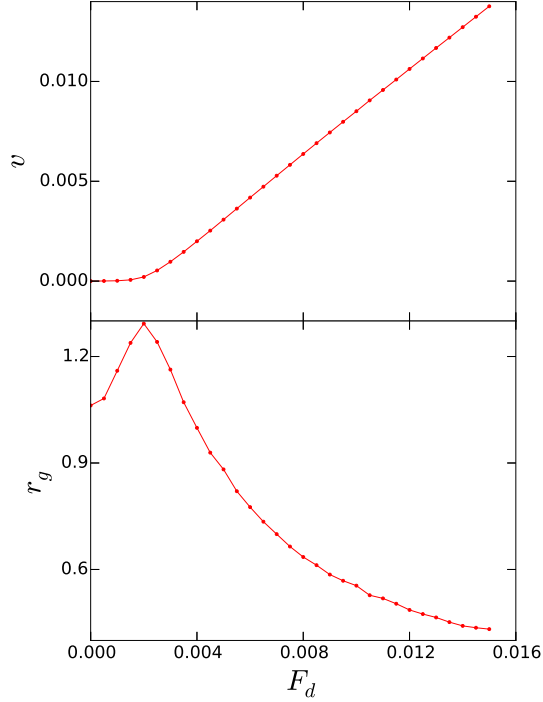} 
\caption{\label{fig1} Mean vortex velocity $v$ (top), in units $b_0 / t_0$, and flux line gyration radius 
$r_g$ (bottom), in units of $b_0$, in the non-equilibrium steady state vs. driving force $F_d$, in units
of $\epsilon_0 / b_0$, for a system of $N = 16$ interacting vortices of length $L = 100\,b_0$ at temperature 
$T = 0.002\,\epsilon_0\,b_0 / k_B$ and for point pinning center strength $p = 0.05\,\epsilon_0$. The top graph 
directly maps onto the current-voltage characteristics for superconductors; the location of the gyration radius
maximum in the bottom graph indicates the depinning threshold \cite{Chaturvedi2015}.}
\end{figure}

When sufficiently large external currents are applied, the vortices detach from the pinning centers and
freely flow through the sample. They then tend to reorder into regular triangular moving lattices. At
zero temperature, the transition from the pinned, glassy phase with drift velocity $v = 0$ to the moving 
state with $v > 0$ is sharp, and represents a continuous non-equilibrium phase transition 
\cite{Fisher1991}. At finite temperatures, this transition becomes smoothened out as the external driving
force $F_d$ is increased. This is shown in Fig.~\ref{fig1} (top panel) for a Langevin molecular dynamics 
simulations of a system of $N = 16$ vortex lines (discretized into $L = 100\,b_0$ layers transverse to the 
magnetic field) driven in a direction perpendicular to the external magnetic field, at temperature 
$T = 0.002\,\epsilon_0 b_0 / k_B \approx 10$ K and with pinning strength $p = 0.05\,\epsilon_0$. Below the 
depinning threshold $F_c \approx 0.002\,\epsilon_0 / b_0$, the flux lines hardly move, but display thermal 
fluctuations; the system remains superconducting. For larger drives, the mean vortex speed scales linearly with
the driving force. This indicates Ohmic dissipation, since $F_d \propto j$ and $v \propto E$, the induced electric
field. Thus the freely flowing vortex system represents a normal-conducting phase.

The Langevin molecular dynamics simulations allow tracking of each individual vortex line, and hence a
detailed analysis of the statistics of the transverse line fluctuations that are induced both thermally
and through interactions with localized pinning sites. At large driving forces $F_d \gg F_c$, the attractions
exerted by point defects become irrelevant, and thermal energies are also minute compared to the work 
associated with the drive. Consequently, the flux lines become straightened out, as is clearly seen in 
Fig.~\ref{fig1} (bottom panel), which displays the mean vortex radius of gyration as function of $F_d$. 
In the disorder-dominated Bragg glass phase, the gyration radius is enhanced by about a factor of three,
indicating the roughness of the vortex trajectories. As $F_d$ approaches $F_c$, the drive pulls more line 
segments away from the attractive defects, which increases the radius of gyration. A distinct maximum of
$r_g$ is observed right at the depinning threshold, a clear signature of the aforementioned zero-temperature 
non-equilibrium phase transition, for which $r_g(F_d \to F_c) \to \infty$ would display a critical divergence
\cite{Dobramysl2013, Chaturvedi2015}.

\begin{figure}[t]
\centering \includegraphics[width=0.94\columnwidth]{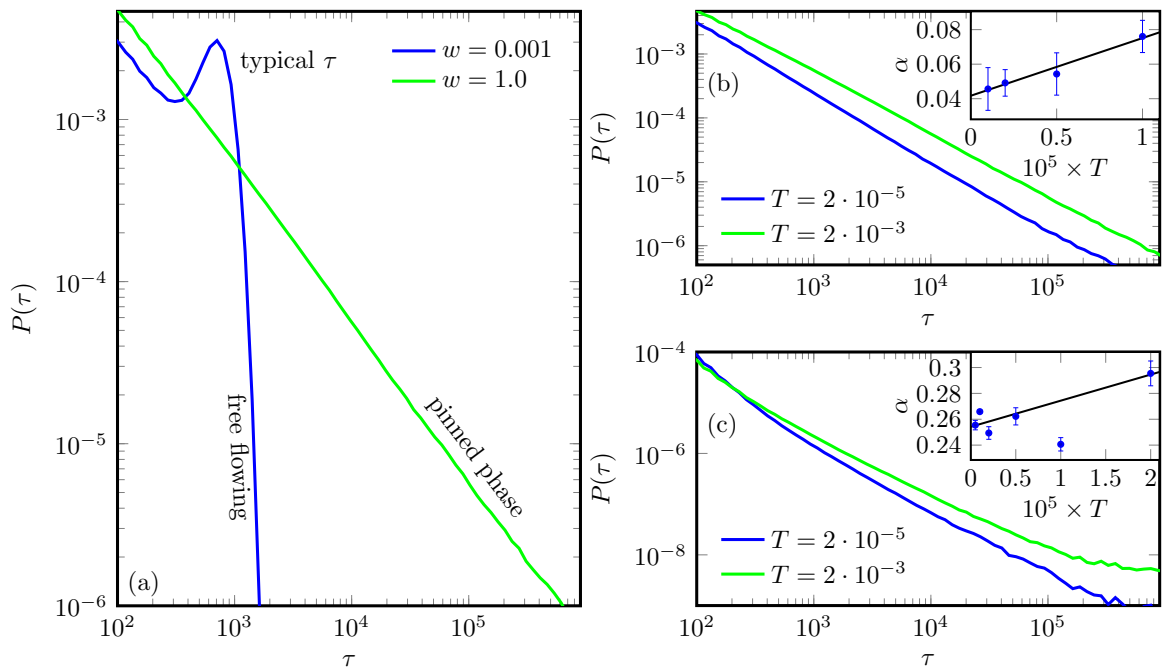}
\caption{\label{fig2} Left panel (a): Distribution of pinning times $P(\tau)$ for single vortices with 
$L / b_0 = 200$ line elements moving through a system with $N_D / L = 100$ randomly placed isolated point 
defects, whose pinning strengths are Gaussian distributed with mean $\mu = 0$ and variance $w^2$; driving 
force $F_d = 0.002\,\epsilon_0 / b_0$, temperature $T = 0.002\,\epsilon_0\,b_0 / k_B$. In the flowing phase 
($w = 0.001\,\epsilon_0$), a characteristic dwell time can be extracted from the observed maximum. In the 
pinned phase ($w = 1.0\,\epsilon_0$), $P(\tau)$ decays algebraically with $\tau$. 
Right panel: (b) Temperature dependence of the power-law scaling and effective decay exponent $\alpha(T)$ for 
the pinning time distribution $P(\tau)$ (with $T$ measured in units of $\epsilon_0\,b_0 / k_B$). (c) Similar
results, but obtained for a continuous Gaussian disorder landscape correlated over distance $\xi = b_0$ with
otherwise identical parameter values \cite{Dobramysl2014}.}
\end{figure}
Even more detailed information on the pinning features of flux lines to variable point disorder is offered by 
careful analysis of the dwelling time statistics. Over many simulation runs, we recorded the durations $\tau$ 
that each line element of a single driven vortex (with $L / b_0 = 200$ segments) spent within a lateral 
distance $b_0$ of its previous position, and gathered the resulting histogram $P(\tau)$ \cite{Dobramysl2014}. 
We kept the driving force $F_d = 0.002\,\epsilon_0 / b_0$ fixed, but varied the temperature $T$ and the width 
(standard deviation) $w$ of an assumed Gaussian distribution for the pinning strengths $p$ about the mean 
$\mu$. Since the point defects are correlated only over a distance $b_0$ in our system, the collective 
Larkin-Ovchinnikov pinning length scale is $L_L \approx b_0 (\tilde{\epsilon}_1 / w)^{2/3}$. The associated 
tpyical energy barrier then becomes $E_L \approx \tilde{\epsilon}_1 b_0^2 / L_L$, from which one estimates the 
critical depinning force as $F_c \approx E_L / (b_0\,L_L) \approx w^{4/3} / (b_0\,\tilde{\epsilon}_1^{1/3})$ 
\cite{Vinokur1996}. Employing extreme-event statistics arguments, Vinokur, Marchetti, and Chen argued that in 
the pinned phase, the pinning time distribution should obey a power law for large dwelling times $\tau$, 
$P(\tau) \sim \tau^{-1-\alpha(T)}$, with an effective temperature-dependent scaling exponent
$\alpha(T) \sim k_B T / E_L$ \cite{Vinokur1996}.

This power-law behavior is indeed borne out by the Langevin molecular dynamics simulation data depicted in
Fig.~\ref{fig2} (left panel, a) at large disorder variance $w = 1.0\,\epsilon_0$. In contrast, in the weak
pinning regime ($w = 0.001\,\epsilon_0$), the vortices remain attached to the defects only for short times
$\tau < 1000\,t_0$, before they detach and freely flow through the system. The marked maximum of $P(\tau)$
observed in the corresponding graph may serve to define a typical pinning time $\tau \approx 700\,t_0$. The
dynamical phase diagram associated with the transition between the freely flowing and the pinned regime has
been studied by Krauth et al by numerically extracting the structure factor exponent of an elastic string in
the $T\to 0$ limit \cite{Kolton2006} and in the finite temperature case \cite{Kolton2009}. The right panel
(b) of the figure explores the temperature dependence of the effective decay exponent $\alpha(T)$; as the
inset for the low-temperature regime $T \leq 10^{-5}\,\epsilon_0\,b_0/k_B$ confirms, one indeed finds a
roughly linear dependence $\alpha(T) \approx 0.0417 + 3340\,T$, albeit with an apparent non-zero intercept as
$T \to 0$ \cite{Dobramysl2014} that is not captured by the scaling theory. We also investigated the dwelling
time distribution where the discrete point defects are replaced by a smooth, continuous disorder landscape, in
order to make more direct contact with the theoretical setting in Ref.~\cite{Vinokur1996}. To this end, we
drew the pinning potential strength $p$ at each node of a square lattice with spacing $b_0$ from a Gaussian
distribution with vanishing mean at variance $w^2$, and constructed the continuous landscape from the
resulting grid points through straightforward bi-linear interpolation. In the pinned state, one again observes
algebraic decay of $P(\tau)$, but actually less cleanly, as shown in the the right panel (c) of
Fig.~\ref{fig2}; also, the typical dwelling time in this situation turns out to be $\tau \approx 100\,t_0$,
indicating a different overall energy scale as compared with the discrete pin case. At low temperatures, we
found the linear relationship $\alpha(T) \approx 0.2542 + 2019\,T$ (inset) \cite{Dobramysl2014}.
 
\subsection{Relaxation dynamics of vortex lines in disordered type-II superconductors}

\begin{figure}[t]
\centering \includegraphics[width=0.50\columnwidth]{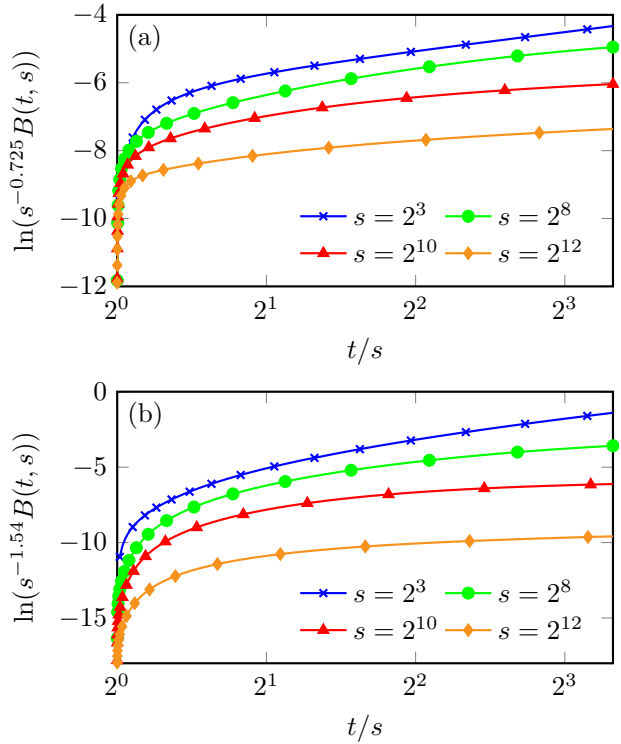}
\caption{\label{fig3} Relaxation of the two-time mean-square displacement $B(t,s)$ for flux lines of length
$L = 640\,b_0$ subject to attractive point defects of strength $p = 0.05\,\epsilon_0$, at temperature
$T = 0.002\,\epsilon_0 b_0 / k_B$, and for various waiting times $s$ as indicated, starting from initially 
straight vortices placed at random locations: (a) in the absence of mutual interactions; (b) in the presence 
of repulsive forces. Simple aging scaling collapse of the data can only be achieved for small waiting times 
$s$, with the aging exponents $b \approx 0.725$ and $b \approx 1.54$. \cite{Dobramysl2013}.}
\end{figure}
Next we proceed to investigate the non-equilibrium relaxation kinetics of vortex matter in disordered type-II 
superconductors starting from various initial configurations that are quite distinct from the stationary states
that may be reached after long time durations. In this brief overview, we report Langevin molecular dynamics 
simulation data for mutually interacting flux lines subject to randomly distributed attractive point pinning 
centers, and probe the relaxation kinetics through the various two-time quantities introduced in section~3.3.  
We first study simulations for systems where initially perfectly straight vortex lines were placed at random 
positions in the sample, and then allowed to relax towards Bragg glass equilibrium states at low temperatures
$T = 0.002\,\epsilon_0 b_0 / k_B \approx 10$K at vanishing driving force $F_d = 0$. We shall attempt to fit 
the data to the simple aging scaling form $A(t,s) \sim s^b \, F_A(t/s)$ for two-time observable $A$ with the 
aging exponent $b$. Indeed, in the absence of disorder and mutual vortex interactions, this setup leads to 
simple aging scaling for both the two-time mean-square displacement $B(t,s)$ and the two-time height-height 
autocorrelation function $C(t,s)$ with the Edwards-Wilkinson scaling exponent $b = 0.5$ \cite{Rothlein2006,
BCI, Dobramysl2013}. 

\begin{figure}[t]
\centering \includegraphics[width=0.50\columnwidth]{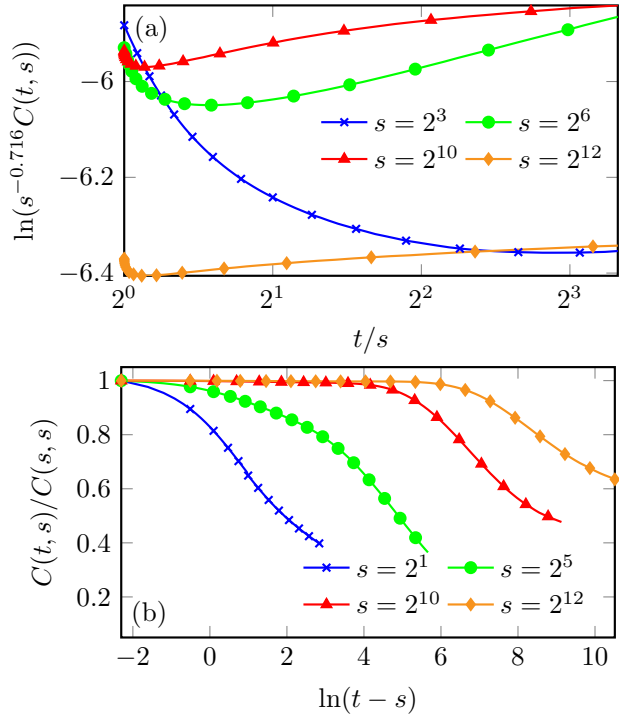}
\caption{\label{fig4} Relaxation of the two-time height-height autocorrelation function $C(t,s)$ for flux lines
of length $L = 640\,b_0$ subject to attractive point defects of strength $p = 0.05\,\epsilon_0$, at temperature
$T = 0.002\,\epsilon_0 b_0 / k_B$, and for various waiting times $s$ as indicated, starting from initially straight
vortices placed at random locations: (a) in the absence of mutual interactions; (b) in the presence of repulsive
forces. \cite{Dobramysl2013}.}
\end{figure}
In the presence of attractive point-like pinning centers, however, such simple aging scaling behavior is 
not observed, although time translation invariance is manifestly broken. Figure~\ref{fig3} displays 
data for the two-time mean-square displacement for various waiting times $s$, both for non-interacting flux 
lines (a) and mutually repelling vortices (b) with $L / b_0 = 640$ segments. Simple aging data collapse can 
only be obtained at short times $s$, with the scaling exponents $b \approx 0.725$ and $b \approx 1.54$, 
respectively. Non-interacting vortices strive to optimize the balance between pinning energy gains and elastic
stretching energy losses. Interacting flux lines in addition become `caged' through the repulsive interactions
with their neighbors, which drastically reduces lateral line fluctuations and displacements. These same
competitive effects are visible also in the simulation data for the height-height autocorrelation function
$C(t,s)$. Simple aging scaling again ensues only for very short waiting times $s$, with, e.g., a scaling
exponent $b \approx 0.716$ for non-interacting vortices as indicated in Fig.~\ref{fig4}(a). The competition
between elastic forces and pinning induce remarkably complex and even non-monotonic relaxation features. When
the repulsive in-plane pair interactions between flux line elements are taken into account, the two-time
height-height autocorrelation data display characteristic two-step relaxation behavior, see Fig.~\ref{fig4}(b),
as is also frequently seen in structural and spin glasses \cite{Gotze1992}. In the `$\beta$' relaxation time 
window, $C(t,s)$ changes hardly at all, and the data appear to satisfy time translation invariance; in the 
ultimate `$\alpha$' relaxation regime, the normalized height-height autocorrelation function finally decays, 
albeit very slowly \cite{Dobramysl2013}. 

We pause for a moment to mention the work of Bustingorry et al \cite{BCD1,BCD2} in which relaxation processes
in a related system were studied using Langevin molecular dynamics. In that work both attractive
and repulsive pinning sites were present in the system. This modification yields important changes
in the relaxation processes. For instance, instead of observing the two-step glassy-like relaxation
shown in Fig. \ref{fig4}, a simple aging scaling of the autocorrelation prevails.

\begin{figure}[t]
\centering \includegraphics[width=0.42\columnwidth]{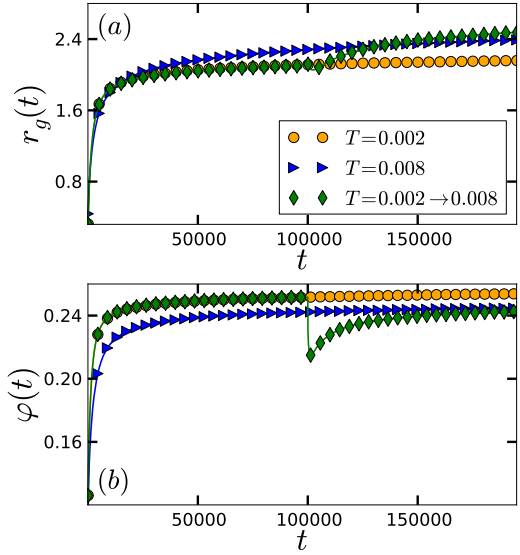}
\caption{\label{fig5} Time evolution of (a) the radius of gyration $r_g(t)$, and (b) the fraction of pinned
vortex line elements $\varphi(t)$, following a sudden temperature quench from $T = 0.002\,\epsilon_0 b_0 / k_B$
to $T = 0.008\,\epsilon_0 b_0 / k_B$ at $t = 100,000\,t_0$ for a system of $N = 16$ interacting flux lines of 
length $L = 640\,b_0$ subject to attractive point defects of strength $p = 0.05\,\epsilon_0$. For comparison,
the time tracks at fixed initial and final temperatures are shown as well. The system relaxes exponentially 
fast to its new steady state \cite{Assi2015}.}
\end{figure}
Experimentally, the initial configuration of randomly placed straight vortices studied above cannot be 
realized. In order to address more realistic initial conditions, we therefore ran a series of numerical
experiments wherein the system is allowed to relax for $10^5$ time steps, until at $t = 100,000\,t_0$ a sudden
change of an external experimental control parameter is implemented. We then monitored several observables to
characterize the system's non-equilibrium relaxation features following such quenches. A first example is
depicted in Fig.~\ref{fig5}: At the quench point, the system's temperature was instantaneously raised from 
$T = 0.002\,\epsilon_0 b_0 / k_B$ to $T = 0.008\,\epsilon_0 b_0 / k_B$. Both the radius of gyration $r_g(t)$
and the fraction of pinned flux line elements $\varphi(t)$, defined as the fraction of segments that reside
within a distance $b_0$ of the attractive point pins at time $t$, are seen to relax essentially exponentially 
towards the system's new steady state at elevated temperature. For the gyration radius, the relaxation time in 
Fig.~\ref{fig5}(a) is measured to be $\tau \approx 34600\,t_0$. Asymptotically, the temporal evolution 
approaches that of an unperturbed vortex sample at the higher temperature. Careful inspection of 
Fig.~\ref{fig5}(a) reveals a tiny dip immediately following the quench before the expected larger gyration
radius value at elevated temperature is reached. Indeed, as confirmed in panel (b), at first a few line 
elements become thermally depinned by the sudden temperature increase, which initially slightly relaxes the 
disorder-induced vortex line roughening in the Bragg glass phase. At any rate, the exponentially fast
relaxation prevents the emergence of a sizeable aging time window \cite{Assi2015}. 

\begin{figure}[t]
\centering \includegraphics[width=0.65\columnwidth]{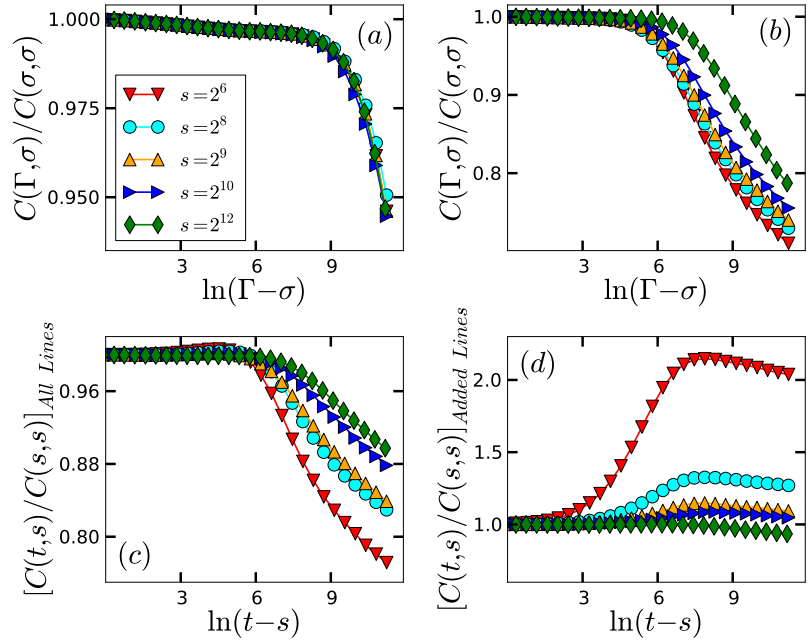}
\caption{\label{fig6} Relaxation of the normalized two-time height-height autocorrelation function for various 
waiting times $s$ (as indicated) at temperature $T = 0.002\,\epsilon_0 b_0 / k_B$ for flux lines of length 
$L = 640\,b_0$ subject to attractive point defects of strength $p = 0.05\,\epsilon_0$ (a) at fixed magnetic field 
or flux density, with $N = 16$ flux lines in our simulation sample; (b) following a field down quench, suddenly 
reducing the vortex number from $N = 21$ to $16$; (c) following a field up quench, where $5$ additional flux lines 
are added to the original $N = 16$ vortices; (d) as in (c), but displaying the data only for the $5$ newly added 
flux lines \cite{Assi2015}.}
\end{figure}
Considerably richer and more complex relaxation features are observed following a magnetic field quench at time
$r = 100,000\,t_0$, when the number of vortices in the system is either suddenly reduced (down quench) or 
increased (up quench), in both cases maintaining a fixed temperature $T = 0.002\,\epsilon_0 b_0 / k_B$ (and a 
vanishing driving force $F_d = 0$). We introduce the total elapsed simulation times $\Gamma = r + t$ and 
$\sigma = r + s$; since $r$ will be much larger than the times $s$ and $t$, both measured after the quench at
$r$, the shift by the duration $r$ needs to be taken into account in attempts to achieve simple aging scaling.
Figure~\ref{fig6}(a) displays the relaxation curves for the normalized two-time height-height autocorrelation 
function with the magnetic field or vortex density held constant for various waiting times $s = \sigma - r$; 
the data essentially correspond to the longest waiting time curve in Fig.~\ref{fig4}(b) and demonstrate that
for the large duration $r = 10^5\,t_0$ considered here, the vortex system has basically reached the stationary 
regime for which time translation invariance holds: the data collapse when plotted against the time difference 
$\Gamma - \sigma = t - s$. In contrast, clear evidence of broken time translation invariance is visible in
Fig.~\ref{fig6}(b), where at $r = 100,000\,t_0$ five randomly selected vortices were instantaneously removed 
from the system. One still observes the $\alpha$-$\beta$ two-step relaxation scenario, yet the system has not
reached stationarity yet, akin to Fig.~\ref{fig4}(b) \cite{Assi2015}.

Qualitatively very similar relaxation features are observed in the system of originally $N = 16$ vortices, when 
at $r = 100,000\,t_0$ five new and initially straight flux lines are introduced at random positions in the 
sample, as seen in Fig.~\ref{fig6}(c). Time translation invariance is again manifestly broken in the late 
$\alpha$ relaxation regime. By separately assessing the relaxation data for the two-time height-height
autocorrelation function of just the newly added flux lines, depicted in Fig.~\ref{fig6}(d), it becomes
apparent that the distinction between Figs.~\ref{fig6}(b) and (c) are caused by the marked non-monotonic 
behavior of the lateral line fluctuation relaxations of the added line population, which typically first show a
strong increase followed by a much slower final decay. Only at very long waiting times $s \geq 2^{12}\,t_0$ do
we observe the monotonic $\alpha$-$\beta$ relaxation features. When the new vortices are inserted into the
sample, they experience strong repulsive forces from the originally present flux lines, which increases their
range and thus facilitates their pinning to point defects. This in turn enhances transverse line fluctuations,
until at $\ln(t - s) \approx 7$ optimal pinning configurations have been reached, whence the final slow decay
of the autocorrelation function ensues \cite{Assi2015}.

\begin{figure}[t]
\centering \includegraphics[width=0.50\columnwidth]{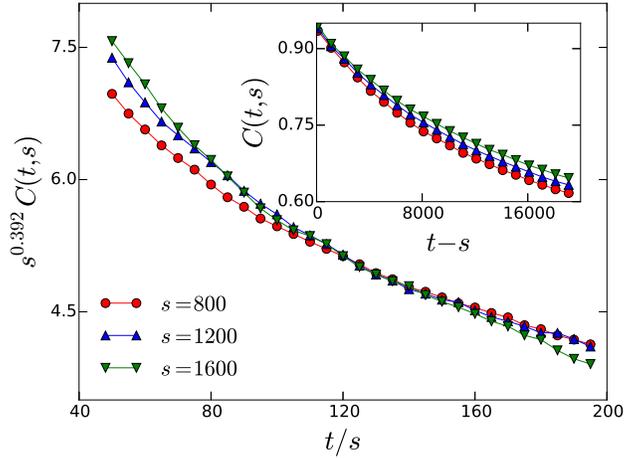}
\caption{\label{fig7} Time evolution of the two-time height-height autocorrelation function for various waiting 
times $s$ (as indicated) at temperature $T = 0.002\,\epsilon_0 b_0 / k_B$ for driven flux lines of length 
$L = 640\,b_0$ subject to attractive point defects of strength $p = 0.05\,\epsilon_0$, as the driving force is
instantaneously reduced from $F_d = 0.015\,\epsilon_0 / b_0$ (i.e., from a moving non-equilibrium steady state) 
into the pinned Bragg glass phase at $F_d = 0.001\,\epsilon_0 / b_0$. The inset shows the data plotted vs. $t-s$,
demonstrating the breaking of time translation invariance; the main panel indicates the best attempt to fit the
data to a simple aging scaling form with $b = 0.392$, which however only partially works in a very limited time
window  \cite{Chaturvedi2015}.}
\end{figure}
As a final set of numerical experiments, we initialize the vortex system with $N = 16$ lines at temperature
$T = 0.002\,\epsilon_0 b_0 / k_B$ and at sufficiently large external driving force 
$F_d = 0.015\,\epsilon_0 / b_0$ to ensure it resides well in the moving non-equilibrium steady state, see 
Fig.~\ref{fig1}. After relaxing the system for a duration of $200,000\,t_0$, the drive is suddenly lowered to
$F_d = 0.001\,\epsilon_0 / b_0$, which forces the flux lines into the pinned, disorder-dominated Bragg glass
phase. Subsequently, we again monitor various observables; as characteristic example, the relaxation data for 
the two-time height-height autocorrelation function $C(t,s)$ are shown in Fig.~\ref{fig7}. As in the previous
magnetic field quench scenarios and for randomized initial conditions, simple aging scaling does not ensue for
this quantity, but holds only approximately in a very limited time regime (with scaling exponent 
$b \approx 0.392$). It is worth mentioning that upon starting in the pinned glassy state, and suddenly 
increasing the driving force into the moving state, the relaxation processes occur exponentially fast and 
quickly reach stationarity, with relaxation times of the order of $1200\,t_0$ \cite{Chaturvedi2015}.

\section{Conclusion}
Langevin molecular dynamics is a powerful numerical method for the investigation of interacting systems
with many degrees of freedom and clear separation of time scales. In this paper we have illustrated this
through the study of interacting vortex matter in different far-from-equilibrium situations. We focused on 
values for the system parameters that are realistic for high-$T_c$ superconductors. When driving the 
vortices through the sample by means of an external current, non-equilibrium steady states
and a (zero temperature) non-equilibrium phase transition emerge. As the Langevin molecular dynamics
simulations allow to track each individual vortex line, a detailed investigation of the steady-state
properties is possible through the measurement of a variety of quantities. A second experimentally
relevant situation arises when a system is brought out of equilibrium through a rapid change (quench) of 
some of the external system control parameters. Vortex matter in a disordered environment provides an 
extremely rich and technologically important model system, with a variety of such perturbations that are
experimentally feasible, namely temperature, magnetic field, as well as current quenches. Detailed
investigations of these different situations allow us to gain a rather complete understanding of the 
relaxation processes that take place when interacting vortex matter is forced out of thermal equilibrium.

\section*{Funding}
This research is supported by the U.S. Department of Energy, Office of Basic 
Energy Sciences, Division of Materials Sciences and Engineering under Award 
DE-FG02-09ER46613.

\end{document}